\documentclass[prl,amssymb,two column]{revtex4}                  
\usepackage{graphicx}
\usepackage{amsmath}
\usepackage{float}
\usepackage{subfigure}
\usepackage{graphicx,epstopdf}

\newcommand{\qlt}{\overline{q_t(L)}}
\newcommand{\qmean}{\overline{q_t}}

\newcommand{\qfinal}{\overline{q_\infty(L)}}

\newcommand{\cm}{\theta_{h}}

\begin{document}
\title{Nature vs.~Nurture: Predictability in Low-Temperature
Ising Dynamics}

\author{J.~Ye}
\email{jingy@princeton.edu} 
\affiliation{
NYU--Courant Institute of Mathematical Sciences,
New York, NY 10012 USA \\
Department of Operations Research and Financial Engineering, Princeton, NJ 08544 USA}
 
\author{J.~Machta}
\email{machta@physics.umass.edu}
\affiliation{
Physics Department,
University of Massachusetts,
Amherst, MA 01003 USA \\
Santa Fe Institute, 1399 Hyde Park Road
Santa Fe, NM 87501 USA}

\author{C.M.~Newman}
\email{newman@cims.nyu.edu}    
\affiliation{
NYU--Courant Institute of Mathematical Sciences, New York, NY 10012 USA \\
Math.~Dept., UC, Irvine, CA 92697 USA}

\author{D.L.~Stein} 
\email{daniel.stein@nyu.edu}
\affiliation{
Physics Department and Courant Institute of Mathematical Sciences,
New York University,
New York, NY 10012 USA}

\begin{abstract}
  Consider a dynamical many-body system with a random initial state
  subsequently evolving through stochastic dynamics. What is the
  relative importance of the initial state (``nature'') vs.~the
  realization of the stochastic dynamics (``nurture'') in predicting
  the final state? We examined this question for the two-dimensional
  Ising ferromagnet following an initial deep quench from $T=\infty$
  to $T=0$. We performed Monte Carlo studies on the overlap between
  ``identical twins'' raised in independent dynamical environments, up
  to size $L=500$ . Our results suggest an overlap decaying with time
  as $t^{-\cm}$ with $\cm = 0.22 \pm 0.02$; the same exponent holds for a quench to low but nonzero temperature. This ``heritability
  exponent'' may equal the persistence exponent for the $2D$~Ising
  ferromagnet, but the two differ more generally. 
\end{abstract}
\maketitle

{\it Introduction.\/}
Given a typical initial configuration of a
thermodynamic system, which then evolves under a specified dynamics,
how much can one predict about what the system will look like at later
times? In this paper, we study this problem in the relatively simple
setting of a homogeneous Ising ferromagnet on a square lattice
following a deep quench from infinite to zero or low temperature. In
particular, we are interested in the influence on
the configuration at a later time $t$ of the random initial state (``nature'') vs.~that 
of the single-spin dynamics (``nurture''), which, even at zero temperature, retains 
an element of randomness in the order that spins are chosen to attempt to flip (and whether they flip when there is no energy change).

This nature vs.~nurture problem~\cite{NS99} can be solved exactly for $1D$ random
ferromagnets and spin glasses~\cite{NNS00}. An initial attempt at a
numerical study of the same problem in the $2D$~homogeneous
ferromagnet on the square lattice was reported in~\cite{ONSS06}; the
results, while suggestive, were inconclusive. In this paper we carry
the earlier studies to a much stronger conclusion concerning
the rate of decay of initial global information in the uniform~$2D$ case. 

The study of nature vs.~nurture
provides a great deal of information on a number of central dynamical issues
concerning different classes of models~\cite{NS99}, such as whether the 
dynamically averaged measure for a single random initial condition settles down
to a limit (even when individual dynamical realizations do not).  Our work is also related to
the general area of phase ordering kinetics~\cite{Mazenko89,Mazenko90,Bray94}.  Several lines
of work are particularly relevant to our study.  Krapivsky, Redner and
collaborators~\cite{SKR118,SKR119,KKR} investigated both the $2D$ and
$3D$~Ising models with zero temperature Glauber dynamics to understand
the time scales and final states of the dynamics.  Derrida, Bray and
Godreche introduced the {\em persistence exponent}~\cite{DBG}, which
characterizes the power law decay of the fraction of spins that are
unchanged from their initial value as a function of time after a
quench.  This exponent was measured for the zero temperature
$2D$~Ising model by Stauffer~\cite{Stauffer94} and calculated exactly for $1D$
by Derrida, Hakim and Pasquier~\cite{DHP95} 
and is studied at nonzero temperature in Refs.~\cite{D97,CS97}.  We will examine the
relationships between the work reported here with these earlier
results.

{\it Heritability.\/} To investigate nature vs.~nurture for the
$2D$~Ising ferromagnet we carry out a ``twin'' study.  We start two
Ising systems with the same infinite temperature initial condition and
then allow each to evolve independently using Glauber dynamics.  We
measure the spin overlap between the systems as a function of time and
refer to this overlap as the ``heritability.''  It is interesting to
note that our notion of heritability is in some sense the opposite of
that in ``damage spreading'' where two slightly different initial
states evolve according to the {\it same\/} dynamical
realization~\cite{Kauffman,Creutz,SSKH,Grassberger,HA98}. 

Our key finding is that heritability, as embodied by the spin overlap
between twins, decays as a power law in time.  For finite systems at $T=0$, an
absorbing state is always reached and we measure the average value of
the overlap between the twins in their final states.  We find that
this quantity decays as a power law in system size.  Based on a finite
size scaling ansatz we obtain a relationship between the power law in
system size for finite systems and the power law in time for infinite
systems.  This relationship is observed to hold in our numerical
results.

{\it Methods.\/} We carried out simulations of the $2D$~ferromagnetic
Ising model on an $L \times L$ square lattice with periodic boundary
conditions.  Each site carries a spin $S_{\rm i} = \pm1, i= 1,2, ...,
N$, where $N = L^2$.  The energy $E$ of the system is
%\begin{equation}
%\label{eq:energy}
$E= -\sum_{<i,j>}S_{i}S_{j}$,
%\end{equation}
where the sum is over nearest neighbor pairs.  

For most of our simulations, the system evolves through zero-temperature Glauber~dynamics.  An
elementary step consists in choosing a random site $i$ and computing
the energy change $\Delta E$ of flipping the spin $S_i$. If $\Delta E
< 0$ the spin is flipped ($S_i \leftarrow -S_i$); if $\Delta E = 0$
the spin is flipped with probability 1/2; and if $\Delta E > 0$ the
spin is not flipped. These moves are repeated until the lattice
reaches a final state where no spin flips are possible. This absorbing
state is either one of the two homogeneous ground states or a striped
state~\cite{SKR118}.  A striped state has one (or more) vertical or
horizontal stripes (but not both) whose boundaries constitute domain
walls separating regions of antiparallel spin orientation.  The
probability that the absorbing state has a single stripe is $0.339\ldots$ for large~$L$, while the probability of multiple
stripes is very small~\cite{SKR118, KKR}.  We also carried out simulations 
of Glauber dynamics at temperature $T=1$, deep in the low temperature phase.  
Here the probability of s spin flipping is proportional to $\exp(-\beta\Delta E)$,
where $\beta$ as usual is the inverse temperature.

To distinguish the influences of nature and nurture, we simulate a
pair of Ising lattices with identical initial conditions (``identical
twins'').  We study the effects of a deep quench, in which the initial
state is at infinite temperature (each spin is independently chosen by
a coin toss).  
The subsequent application of Glauber
dynamics  then effects an instantaneous quench to low or zero temperature.  Each twin evolves {\em independently\/} according to
Glauber dynamics at a fixed temperature, either $T=0$ or $T=1$. For zero-temperature Glauber dynamics, each run is continued until an absorbing
state is reached. The time $t$ is measured in sweeps with one sweep corresponding
to $N$ spin-flip attempts. At each time step $t$, we study the overlap
$q_t(L)= \frac{1}{N}\sum_{i=1}^{N}S_{i}^1(t)S_{i}^2(t)$ between the
twins, where $S_{i}^1(t)$ denotes the state of the $i^{\rm th}$ spin
at time $t$ in twin~1, and similarly for $S_{i}^2(t)$.  The influence
of initial conditions is quantified by $q_t(L)$, where $q_0(L)=1$ for
any~$L$.  We are interested in the average $\overline{q_t(L)}$ over
both initial conditions and the subsequent dynamics and, in
particular, in understanding both the size dependence of final
overlap, $\qfinal = \lim_{t \rightarrow \infty} \overline{q_t(L)}$ and
the time dependence of the infinite volume limit $\qmean = \lim_{L
  \rightarrow \infty} \overline{q_t(L)}$.  As we shall see, the
behavior of $\qfinal$ and $\qmean$ are connected by a finite size
scaling ansatz.

For $T=0$, we studied 21 lattice sizes from $L=10$ to $L=500$.  For each size we
studied 30,000~independent pairs of twins out to a time such that almost
all systems are in an absorbing state.  For each initial condition we
compute only two dynamical trajectories, one for each twin.  This
approach is statistically equivalent and more efficient than averaging
over both the dynamics and initial conditions.  From $q_t(L)$ for each
pair of twins, we computed the mean $\qlt$.

\begin{figure}[h]
\centering
\includegraphics[scale=1]{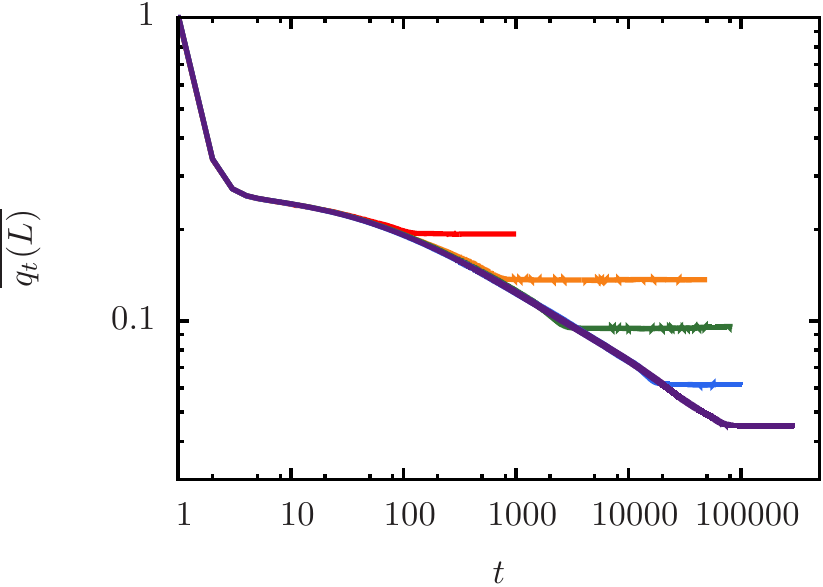}
\caption{(Color online) $\qlt$\ vs.\ $t$ for several $L$ and for quenches to $T=0$.  The  plateau value decreases from small to large $L$. From top to bottom the sizes are $L=20$, 50, 100, 250, and 500.}
\label{fig:MTF}
\end{figure}

{\it Results.\/} Figure~\ref{fig:MTF} shows a log-log plot of $\qlt$
vs.~$t$ for several~$L$ and for quenches to $T=0$ .
We observe that for short and intermediate
times, $\qlt$ appears to follow a single curve for all~$L$, until an
$L$-dependent time scale when $\qlt$ separates from the main curve and
a plateau is reached.  It is reasonable to suppose that the single
curve represents the infinite volume behavior~$\qmean$ to good
approximation.  The initial decay of $\qmean$ is rapid, followed by a
shoulder that goes to about $t=100$.  The subsequent behavior appears
to be described by a power law.  A power law fit of the form $\qmean=d
t^{-\cm}$ for the largest two sizes, $L=400$ and $L=500$, from $t=500$
to $t=10^4$ yields $d = 0.59(3)$ and $\cm = 0.216(7)$ for $L=400$,
and $d = 0.62(3)$ and $\cm= 0.225(6)$ for $L=500$.  The error bars
are obtained by the bootstrap method.  Based on these two sizes, we
estimate that the heritability exponent describing the decay of the
overlap with time is $\cm=0.22\pm 0.02$.

Next we discuss $\qfinal$, the finite size behavior of the absorbing
value of $\qlt$ for quenches to $T=0$. For $L < 300$ we simulate all systems until they are
absorbed, but for the largest several sizes this is not possible.
Therefore, for the small fraction of pairs with at least one
unabsorbed twin, we approximate their contribution to $\qfinal$ by the
value of the $q_t(L)$ at the largest $t$, which is on the plateau for
all sizes.  The justification for this approximation is that the
plateau value of $\qlt$ is nearly equal to $\qfinal$ for two
reasons. First, most twins are absorbed before the plateau is
reached. The second reason is more subtle.  On the plateau essentially
all unabsorbed systems are in a diagonal stripe state.  The diagonal
stripe, appearing in about 4\% of systems, is the longest lived
metastable state with a lifetime that scales as $L^3$~\cite{SKR118}.
A diagonal stripe decays by a random walk in the width of the stripe
and almost always reaches one of the two homogeneous ground states.  A
simple random walk argument shows that the probability of the final
state being all $+1$ is equal to the fraction of $+1$ spins in the
diagonal stripe state.  Thus the expected value of $\qfinal$ averaged
over the dynamics is approximately equal to $\qlt$ on the plateau,
which justifies the approximation.  For the largest size studied
($L=500$), approximately 8\% of pairs of twins are not yet absorbed at
the end of the simulation.  This fraction is to be expected if all
unabsorbed pairs have a twin in a diagonal stripe state.  Comparison
between this approximation and the exact simulation for smaller sizes
shows that the error due to the approximation is much smaller than the
statistical errors due to the finite sample size.

\begin{figure}[h]
 \centering
\includegraphics[scale=1]{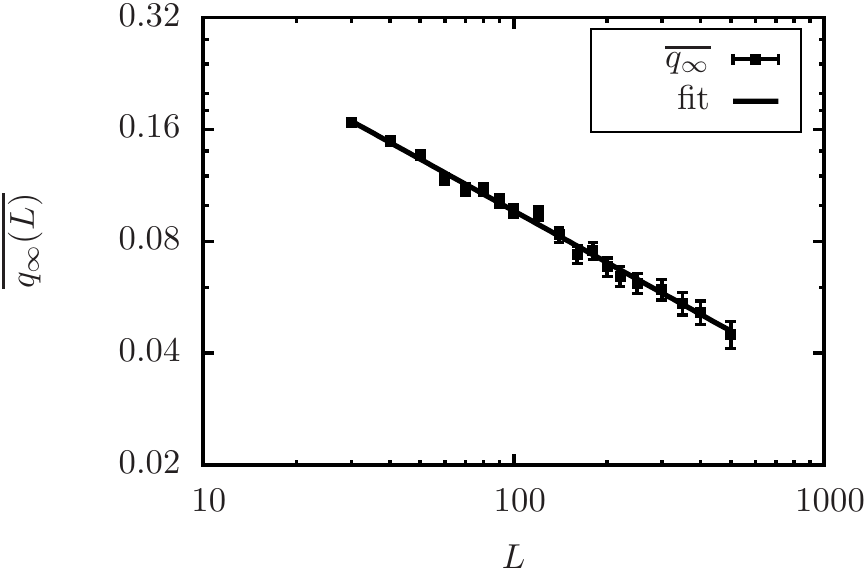}
\caption{$\qfinal$\ vs.\ $L$ for quenches to $T=0$. The solid line is the best power law fit
  for sizes 20 to 500, and corresponds to $\qfinal\sim L^{-0.46}$.}
\label{fig:MSF}
\end{figure}

The results for $\qfinal$ for quenches to $T=0$ are shown in Fig.~\ref{fig:MSF}.  We find
that the data are well fit by a power law of the form $\qfinal = a
L^{-b}$.  We performed a series of power law fits in which we
successively dropped smaller sizes.  The only significant deviation
from a power law fit occurs if the data from the smallest size, $L=10$,
are included. The quality of the fits is good and the results
independent of the minimum size within the error bars indicating that
the corrections to scaling are small.  For example, if $L=10$ and 20
are excluded the best fit result is $a= 0.81(3)$, $b=0.46(1)$ with
$\chi^2/{\rm d.o.f.} =0.51$ and the corresponding quality of fit is
$Q=0.95$.  Our best estimate of $b$, taking into account both
statistical and possible systematic errors, is $0.46 \pm 0.02$.

The exponents $b$ and $\cm$ can be related via a finite-size scaling
ansatz.  During coarsening, the typical domain size $\xi$ grows as
$\xi \sim t^{1/z}$ where $z$ is the dynamic exponent for coarsening
and $z=2$ for zero temperature Glauber dynamics~\cite{Bray94}.  We
therefore postulate the following finite size scaling form for $\qlt
\sim t^{-\cm}f(\frac{t^{1/z}}{L})$, where the function $f(x)$ is
expected to behave as:
\begin{equation}
\label{eq:func}
f(x) \sim \begin{cases}
1\ \ \ \   \text{for}\ x\ll1 , \\
x^{z \cm}\ \ \text{for}\ x\gg 1.
\end{cases}
\end{equation}
The large $x$ behavior is required to ensure that $\qlt$ approaches a
constant for large $t$ and finite $L$.  In particular, the $t \to
\infty$ behavior is $\qfinal \sim L^{-z \cm}$, so that $b = z \cm=2
\cm$, which agrees within errors with our numerical results $b=0.46\pm
0.02$ and $\cm=0.22 \pm 0.02$. Since $b$ is obtained from the data for
many sizes while $\cm$ is obtained from a limited range in $t$ and
only two sizes, we consider $b$ to be a more reliable value.  We also
considered initial conditions with random spins but fixed
magnetization rather than infinite temperature.  If the fixed
magnetization is within O($1/\sqrt{N}$) of zero, the results for $b$ are
the same as for the infinite temperature case.

\begin{figure}[h]
 \centering
\includegraphics[scale=1]{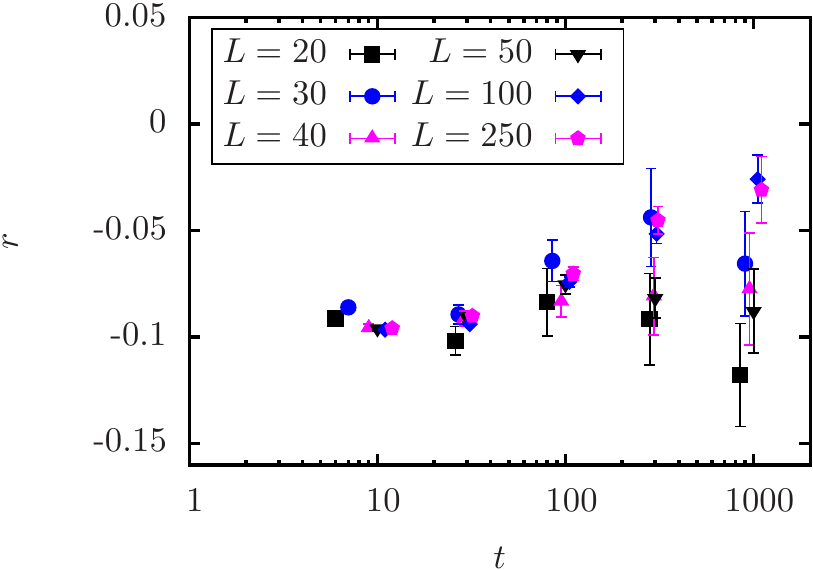}
\caption{(Color online) The fractional difference between the $T=1$ and $T=0$ heritability, $r(t)$ vs.\ time $t$ for several system sizes $L$.}
\label{fig:rt}
\end{figure}

Next we consider quenches to $T=1$.  These simulations are more difficult to carry out than for $T=0$ and we use kinetic Monte Carlo to accelerate them.  The curves for $\qlt$ vs.~$t$ fall slightly below those found for $T=0$ (see Fig.\ \ref{fig:MTF}) as would be expected since thermal fluctuations should decrease $\qlt$ relative to zero-temperature dynamics.  Figure \ref{fig:rt} shows the ratio $r(t)=(\qlt_{T=1}-\qlt_{T=0})/\qlt_{T=0}$ for various sizes.  While the error bars are large, we see that $r(t)$ is approximately constant as a function of both $t$ and $L$ suggesting that the same heritability exponent applies for both $T=0$ and $T=1$.  We believe that similar results would hold throughout the low temperature phase.  It would be interesting to study quenches to the critical temperature where one might expect a different heritability exponent. 

{\it Heritability and persistence.\/} Heritability is related, at
least superficially, to the phenomena of persistence.  In the context
of phase ordering kinetics, persistence is defined as the fraction of
spins that have not flipped  from their initial values up to time $t$. This
quantity is found to decay as a power law and the exponent $\theta_p$
is called the ``persistence exponent.''  Numerical simulations on the
$2D$~Ising model with zero temperature non-conserved dynamics
yield~\cite{Stauffer94} $\theta_p=0.22$ (with no error bars quoted)
and~\cite{J99} $\theta_p=0.209(2)$ (error bars are only statistical).
These numbers are within the error bars of the heritability exponent
$\cm=0.22\pm0.02$ obtained here.  

In addition, our exponent $b=0.46\pm 0.02$ describing the finite size
decay of heritability can be compared directly to the finite-size
persistence exponent $\theta_{\rm Ising}=0.45\pm0.01$~\cite{MS96}.
(As discussed in~\cite{MR00}, the same finite scaling arguments that
show that $b = z \cm$ demonstrate that $\theta_p$ defined
in~\cite{Stauffer94} and $\theta_{\rm Ising}$ defined in~\cite{MS96}
are related by $\theta_{\rm Ising}= z \theta_p$.)  

In contrast, for the one-dimensional ferromagnetic Ising model one can
compute analytically both the persistence exponent and the
heritability exponent using the mapping from zero-temperature Glauber
dynamics to the voter model and to coalescing random walks (see,
e.g.,~\cite{DHP96, FINS01}).  It is shown in~\cite{DHP95, DHP96} that
$\theta_p=3/8$, but related arguments can be used to show that $\cm =
1/2$ and $b = 1$.  While the persistence and heritability exponents
are distinct in one dimension, it may be that they are exactly the
same in two dimensions or it may be that they are simply close but not
identical.

{\it Discussion.\/} 
Motivated by general questions concerning the
predictive power of initial configurations in systems evolving by stochastic dynamics,
we studied the simple $2D$~homogeneous Ising ferromagnet with an
initial state quenched from $T=\infty$ to $T=0$ (and $T=1$). In analogy to the
classic approach to nature (initial configuration) vs.~nurture
(dynamical realization), we performed Monte Carlo studies of identical twins,
$S^1(t)$ and $S^2(t)$, raised in independent dynamical environments
with system sizes up to $L=500$.

The quantity we focused on was the overlap $q_t(L)$ between $S^1(t)$
and $S^2(t)$, which was then averaged over 30,000 sets of identical
twins to give $\qlt$. Extensive studies of the asymptotic
behavior of $\qlt$ for large $t$ and $L$ were performed.  Our
main conclusions are as follows:
\begin{itemize}
  \item 
  For finite $L$ and $T=0$, there are limiting absorbing states $S^j(\infty)$
and overlaps $\qfinal  \sim a L^{-b}$ with $b = 0.46 \pm 0.02$.

  \item 
For zero and low temperature  $\qlt$ appears to approach an infinite volume limit $\qmean$
as $L \to \infty$ with $\qmean \sim d t^{-\cm}$ and $\cm=0.22 \pm 0.02$.  For the 2D Ising ferromagnet $\cm >0$ so that memory of the initial state eventually decays to zero. However, the small value of $\cm$ shows that information about the initial state decays rather slowly.

  \item 
  Finite size scaling considerations suggest that $b=2\cm$, consistent
with our numerically estimated values and with the exact $1D$ values.

\item The numerical values of the heritability and persistence
  exponents are very close for the $2D$~Ising model, though these
  exponents are distinct in one dimension: for the $1D$~Ising model
  the heritability exponent is larger than the persistence exponent.

\end{itemize}

%It would be interesting to study heritability both in higher dimension
%and in disordered systems and to understand the relationship between
%persistance and heritability in these situations.
It is an interesting open question whether the heritability exponent is an independent non-equilibrium exponent or whether it is related to either the persistence exponent or, as suggested by an anonymous referee, the autocorrelation exponent.  Additional simulations either of the Ising model in higher dimensions or for Potts models may resolve this question.

\begin{acknowledgements} 
  The research of JM was supported in part by NSF DMR-1208046.  
  %The research of CMN and DLS was supported in part by NSF DMS-1207678.
  The research of JY, CMN and DLS was supported in part by NSF DMS-1207678
and that of CMN and DLS also in part by NSF OISE-0730136.
  Simulations were performed on the Courant Institute of Mathematical
  Sciences computer cluster and the University of Massachusetts
  Condensed Matter Theory cluster.  We thank Sidney~Redner for
  interesting discussions.
\end{acknowledgements}

\small\def\em{\it} \newcommand{\noopsort}[1]{} \newcommand{\printfirst}[2]{#1}
  \newcommand{\singleletter}[1]{#1} \newcommand{\switchargs}[2]{#2#1}

\end{document}